\documentclass{iopart}
\usepackage{amsmath,amssymb}
\usepackage{tikz}
\usetikzlibrary{shapes,positioning}

\DeclareMathAccent{\wtilde}{\mathord}{largesymbols}{"65}
\DeclareMathAccent{\what}{\mathord}{largesymbols}{"62}

\def\m@th{\mathsurround=0pt}
\mathchardef\bracell="0365 
\def\upbrall{$\m@th\bracell$}
\def\undertilde#1{\mathop{\vtop{\ialign{##\crcr
    $\hfil\displaystyle{#1}\hfil$\crcr
     \noalign
     {\kern1.5pt\nointerlineskip}
     \upbrall\crcr\noalign{\kern1pt
   }}}}\limits}
\def\theequation{\arabic{section}.\arabic{equation}}

\newcommand{\wh}{\widehat}
\newcommand{\wt}{\widetilde}

\newcommand{\ol}{\overline}

\newcommand{\al}{\alpha}

\newcommand{\gm}{\gamma}

\newcommand{\dl}{\delta}
\newcommand{\Dl}{\Delta}
\newcommand{\ep}{\epsilon}

\newcommand{\lm}{\lambda}
\newcommand{\Lm}{\Lambda}

\newcommand{\vf}{\varphi}

\newcommand{\cL}{\mathcal{L}}

\newcommand{\bblu}{\begin{color}{blue}}
\newcommand{\bred}{\begin{color}{red}}
\newcommand{\ecl}{\end{color}}

\newcommand{\nn}{\nonumber}

\newcommand{\be}{\begin{equation}}
\newcommand{\ee}{\end{equation}}
\newcommand{\bea}{\begin{eqnarray}}
\newcommand{\eea}{\end{eqnarray}}
\newcommand{\bse}{\begin{subequations}}
\newcommand{\ese}{\end{subequations}}

\begin{document}
\title{Lagrangian multiform structure for the lattice Gel'fand-Dikii hierarchy} 
\author{S.B. Lobb and F.W. Nijhoff}
\address{Department of Applied Mathematics, University of Leeds, Leeds LS2 9JT, UK}

\begin{abstract}
The lattice Gel'fand-Dikii hierarchy was introduced in \cite{NijPapCapQui1992} as the family of partial difference equations generalizing to higher rank the lattice Korteweg-de Vries systems, and includes in particular the lattice Boussinesq system. We present a Lagrangian for the generic member of the lattice Gel'fand-Dikii hierarchy, and show that it can be considered as a Lagrangian 2-form when embedded in a higher dimensional lattice, obeying a closure relation. Thus the multiform structure proposed in \cite{LobNij2009} is extended to a multi-component system.
\end{abstract}

\section{Introduction}
Multidimensional consistency \cite{NijWal2001,BobSur2002} has come to be regarded as one of the hallmarks of integrability for discrete systems. In brief, it is the property that several copies of an equation may be imposed simultaneously on a higher dimensional lattice, and no inconsistency or multivaluedness occurs in the evaluation of the dependent variables at each lattice site. Lagrangian multiforms were proposed for multidimensionally consistent discrete systems in \cite{LobNij2009} to remedy the fact that while several equations may coexist on the multidimensional lattice, only one equation can be derived from a scalar Lagrangian. The multiform structure allows copies of the relevant equation in all possible lattice directions to be derived through the Euler-Lagrange equations, making multidimensional consistency manifest on the level of the Lagrangian.

The lattice systems considered in the first instance were quadrilateral equations of the form
\be
 Q(u,u_{\mu},u_{\nu},u_{\mu\nu};\al_{\mu},\al_{\nu}) = 0,
\ee
where $u=u(n_{\mu},n_{\nu})$ depends on two discrete variables $n_{\mu},n_{\nu}$, shifts of $u$ in the $n_{\mu}$-direction are denoted by $u_{\mu}$ (so that for example $u_{\mu}=u(n_{\mu}+1,n_{\nu})$), and the $\al_{\mu}$ are lattice parameters associated with the $n_{\mu}$-direction. The examples, defined on elementary plaquettes in a 2-dimensional lattice, were chosen from the classification given in \cite{AdlBobSur2003}.

The important observation was that all of the systems considered in \cite{LobNij2009} admit Lagrangians $\cL_{\mu\nu}=\cL_{\mu\nu}(u,u_{\mu},u_{\nu};\al_{\mu},\al_{\nu})$ in terms of 3 points, which satisfy the following closure relation
\be\label{2dclos}
 \Dl_{\lm}\cL_{\mu\nu}+\Dl_{\mu}\cL_{\nu\lm}+\Dl_{\nu}\cL_{\lm\mu} = 0,
\ee
where the difference operator $\Dl_{\lm}$ acts on functions $f$ of $u=u(n_{\lm},n_{\mu},n_{\nu})$ by the formula $\Dl_{\lm}f(u)=f(u_{\lm})-f(u)$, and on a function $g$ of $u$ and its shifts by the formula $\Dl_{\lm}g(u,u_{\mu},u_{\nu})=g(u_{\lm},u_{\lm\mu},u_{\lm\nu})-g(u,u_{\mu},u_{\nu})$. This allowed the interpretation of the Lagrangian as a closed 2-form on the multidimensional lattice, and on the basis of this a new variational principle was proposed.

This idea was extended in \cite{LobNijQui2009} to a 3-dimensional system, the bilinear discrete Kadomtsev-Petviashvili (KP) equation. This is an equation in 6 points, with a Lagrangian $\cL_{\lm\mu\nu}$ also in terms of 6 points. In this case the relevant closure relation obeyed by the Lagrangian is
\be\label{3dclos}
 \Dl_{\rho}\cL_{\lm\mu\nu}-\Dl_{\lm}\cL_{\mu\nu\rho}+\Dl_{\mu}\cL_{\nu\rho\lm}-\Dl_{\nu}\cL_{\rho\lm\mu} = 0,
\ee
where as before the difference operator $\Dl_{\lm}$ acts on functions $f$ of $\tau=\tau(n_{\lm},n_{\mu},n_{\nu},n_{\rho})$ by the formula $\Dl_{\lm}f(\tau)=f(\tau_{\lm})-f(\tau)$, and on a function $g$ of $\tau$ and its shifts by the formula $\Dl_{\lm}g(\tau,\tau_{\mu},\tau_{\nu},\tau_{\rho})=g(\tau_{\lm},\tau_{\lm\mu},\tau_{\lm\nu},\tau_{\lm\rho})-g(\tau,\tau_{\mu},\tau_{\nu},\tau_{\rho})$. Here, the Lagrangian can be considered as a closed 3-form.

So far, all systems with Lagrangians that have been shown to obey a closure relation have involved variables around a simple plaquette or cube. This is also the case for the lowest order member in what is effectively a lattice analogue of the Gel'fand-Dikii (GD) hierarchy \cite{NijPapCapQui1992}, the discrete Korteweg-de Vries (KdV) equation, which was one of the systems considered in \cite{LobNij2009}.
The next system in the hierarchy is the discrete Boussinesq equation \cite{NijPapCapQui1992}, this involves not only nearest neighbouring points, but also next-nearest neighbours. Along with higher order systems in the GD hierarchy, it can be written as a coupled system of partial difference equations defined on an elementary plaquette. In this paper we present a Lagrangian for the generic member of the lattice GD hierarchy, and show that it obeys the closure relation \eqref{2dclos}, allowing us to apply the variational principle proposed in \cite{LobNij2009}.

\section{The lattice Gel'fand-Dikii hierarchy}
The lattice GD hierarchy first appeared in \cite{NijPapCapQui1992}, where the direct linearization method was used to find a discrete analogue of the continuous GD hierarchy, which is a hierarchy of systems associated with higher order spectral problems \cite{GelDik1976,GelDik1979,Man1979,DriSok1985}. As already mentioned, the first members in the hierarchy are the lattice KdV and lattice Boussinesq equations, higher order members are coupled systems of partial difference equations in terms of variables $u,v_{j},w_{j}$, where $0\leq j\leq N-2$, given by the following. 
\bse
\be\label{v}
 \wh{v}_{j+1}-\wt{v}_{j+1} = (p-q+\wh{u}-\wt{u})\wh{\wt{v}}_{j}-p\wh{v}_{j}+q\wt{v}_{j}
\ee
\be\label{w}
 \wh{w}_{j+1}-\wt{w}_{j+1} = -(p-q+\wh{u}-\wt{u})w_{j}-q\wh{w}_{j}+p\wt{w}_{j}
\ee
for $0\leq j\leq N-3$, and
\bea
\fl (p-q+\wh{u}-\wt{u})(\wh{\wt{v}}_{N-2}-w_{N-2}) && =
       (p+q+u)\bigl[(p-q+\wh{u}-\wt{u})\wh{\wt{v}}_{N-3}-p\wh{v}_{N-3}+q\wt{v}_{N-3}\bigr]\nn\\
\fl && +\sum_{j=0}^{N-3}\bigl[(-p)^{N-1-j}(\wt{v}_{j}-w_{j})-(-q)^{N-1-j}(\wh{v}_{j}-w_{j})\nn\\
\fl && -w_{j}\bigl((-p)^{N-2-j}\wt{u}-(-q)^{N-2-j}\wh{u}\bigr)\bigr]\nn\\
\fl && -\sum_{j=2}^{N-2}\sum_{i=0}^{N-1-j}w_{i}\bigl[(-p)^{N-1-j-i}\wt{v}_{j-1}-(-q)^{N-1-j-i}\wh{v}_{j-1}\bigr],\nn\\
\fl && \label{N-2}
\eea
\ese
identifying $v_{0}-w_{0}=u$.
Here we have used the notation we find most instructive: the dependent variables are $u,v_{j},w_{j}$ for $0\leq j\leq N-2$, and we consider them to depend on two independent variables $n,m$. The symbol $\;\wt{}\;$ is used to denote shifts in the $n$-direction, and $\;\wh{}\;$ denotes shifts in the $m$-direction, so that if $u=u(n,m)$, then $\wt{u} = u(n+1,m)$ and $\wh{u}=u(n,m+1)$. The lattice parameters $p,q$ are associated with the $n,m$-directions respectively. This is illustrated in the diagram below.

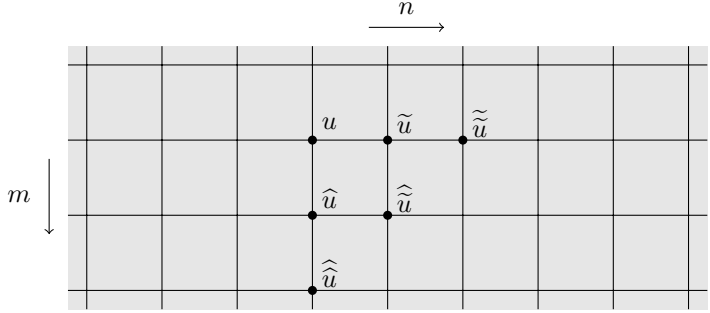
\begin{tikzpicture}[every node/.style={minimum size=1cm},on grid]
 \fill[gray!20] (0.75,0.75) rectangle (9.25,4.25);
 \draw (0.75,0.75) grid (9.25,4.25);
 \fill (4,3) circle (0.06) node[above right=-0.4] {$u$};
 \fill (5,3) circle (0.06) node[above right=-0.4] {$\wt{u}$};
 \fill (6,3) circle (0.06) node[above right=-0.4] {$\wt{\wt{u}}$};
 \fill (4,2) circle (0.06) node[above right=-0.4] {$\wh{u}$};
 \fill (5,2) circle (0.06) node[above right=-0.4] {$\wh{\wt{u}}$};
 \fill (4,1) circle (0.06) node[above right=-0.4] {$\wh{\wh{u}}$};
 \draw[->] (4.75,4.5) -- (5.75,4.5);  \node at (5.25,4.75) {$n$};
 \draw[->] (0.5,2.75) -- (0.5,1.75); \node at (0.1,2.25) {$m$};
\end{tikzpicture}

The variables $v_{j},w_{j}$ are evaluated at the same lattice point as $u$, while $\wt{v}_{j},\wt{w}_{j}$ are evaluated at the same lattice point as $\wt{u}$, and so on.

As noted in \cite{NijPapCapQui1992} the lattice GD hierarchy arises from a Zakharov-Shabat type of linear problem
\begin{equation}
(p+\omega k) \wt{{\bf \vf}}_k\ =\ L_k \cdot {\bf \vf}_k\  ,  \ 
(q+\omega k) \wh{{\bf \vf}}_k\ =\ M_k \cdot {\bf \vf}_k\  , 
\label{eq:8} 
\end{equation}
in which 
\begin{equation}
L_k\ =\ \left( \begin{array}{ccccc}
p-\wt{u}  & 1 &        &           &  \\
-\wt{v}_1 & p & 1      &           &  \\
\vdots            &   & \ddots & \ddots    &  \\
-\wt{v}_{N-2}& 0 & \cdots & p         & 1 \\
k^N+\ast    & w_{N-2}& \cdots & w_1& p+u \\
\end{array} \right) , \label{eq:9}
\end{equation}
and where the matrix $M_k$ is a similar matrix obtained after 
the  replacements $p\mapsto q$ and $\;\wt{}\mapsto \wh{}\;$. 
The term $\ast$ in the left-lower corner of the matrix $L_k$ is 
such that the determinant $\det(L_k)=p^N-(-k)^N$, 
i.e. we have the expression
\be
\ast\ =\ \sum_{j=0}^{N-2}\,(-p)^{N-1-j}\left( \wt{v}_j\,-\,
w_j\right) \ -\ \sum_{j=1}^{N-1}\sum_{i=0}^{N-1-j}\,
(-p)^{N-1-j-i}w_i\tilde{v}_{j-1}\   .
\ee 

Because the system is multidimensionally consistent, a concept introduced independently in \cite{NijWal2001} and \cite{BobSur2002}, we are free to impose copies of the equations in other lattice directions, with appropriate lattice parameters. In particular, suppose we have another lattice direction associated with parameter $r$, where shifts in this direction are denoted by $\;\bar{}\;$. Then copies of the equation \eqref{v} can be imposed on each pair of lattice directions, giving

\bse
\be\label{vpq}
 \wh{v}_{j+1}-\wt{v}_{j+1} = (p-q+\wh{u}-\wt{u})\wh{\wt{v}}_{j}-p\wh{v}_{j}+q\wt{v}_{j}
\ee
\be\label{vqr}
 \ol{v}_{j+1}-\wh{v}_{j+1} = (q-r+\ol{u}-\wh{u})\wh{\ol{v}}_{j}-q\ol{v}_{j}+r\wh{v}_{j}
\ee
\be\label{vrp}
 \wt{v}_{j+1}-\ol{v}_{j+1} = (r-p+\wt{u}-\ol{u})\wt{\ol{v}}_{j}-r\wt{v}_{j}+p\ol{v}_{j}
\ee
\ese
for $0\leq j \leq N-3$.

In particular, summing equations \eqref{vpq},\eqref{vqr} and \eqref{vrp} for $j=0$ gives 
\be\label{LKP}
 0=(p-q+\wh{u}-\wt{u})(\wh{\wt{u}}+r)+(q-r+\ol{u}-\wh{u})(\wh{\ol{u}}+p)+(r-p+\wt{u}-\ol{u})(\wt{\ol{u}}+q),
\ee
which is in fact the lattice Kadomtsev-Petviashvili (or lattice KP) equation \cite{NijCapWieQui1984}. That \eqref{LKP} holds is natural, since members of the lattice GD hierarchy can be viewed as a special type of periodic reduction of the lattice KP equation.

\section{Lagrangian for the lattice GD hierarchy}
Lagrangians for the two lowest order members of the lattice GD hierarchy have already appeared in the literature, an action for the KdV lattice was first given in \cite{CapNijPap1991}, and for the Boussinesq lattice in \cite{NijPapCapQui1992} (these systems arise from the lattice GD hierarchy by taking $N=2$ or $N=3$ respectively). In fact, as we show here, it is possible to write a Lagrangian for the generic member of the hierarchy.

Note first that if we define 
\be
 \gm_{j}(p,q) \equiv \frac{(-p)^{j+1}-(-q)^{j+1}}{-p+q} = (-1)^{j}(p^{j}+p^{j-1}q+\dots+pq^{j-1}+q^{j}),
\ee
then equation \eqref{N-2} can be written in a more convenient form as
\bea 
\fl \gm_{N-1}(p,q)\,+\,\wh{\wt{v}}_{N-2}\,-\,w_{N-2} &=&
\frac{(p-q)\gm_{N-1}(p,q)}{p-q+\wh{u}-\wt{u}}\,+\, 
\sum_{i=0}^{N-3}\sum_{j=0}^{N-3-i} \gm_{N-3-i-j} \wh{\wt{v}}_j w_i \nn \\ 
\fl &&-\,\sum_{j=0}^{N-3}\,\gm_{N-2-j}\left( \wh{\wt{v}}_j\,-\,w_j \right),
\label{N-2_better} 
\eea 
which enables us more easily to see that the following proposition holds.

\paragraph{}
\noindent
\textbf{Proposition 1:} \emph{The system consisting of equations \eqref{v},\eqref{w} and \eqref{N-2_better} solves the discrete Euler-Lagrange equations for the following Lagrangian
\bea
\fl \cL_{pq} & \equiv & (p-q)\gm_{N-1}(p,q)\ln(p-q+\wh{u}-\wt{u})-\gm_{N-1}(p,q)(\wh{u}-\wt{u})\nn\\
\fl                  && \;\;\;\;-\sum_{j=0}^{N-2}\gm_{N-2-j}(p,q)(\wh{u}-\wt{u})\wh{\wt{v}}_{j}\nn\\
\fl                  && -\sum_{i=0}^{N-3}\sum_{j=1}^{N-2-i}\gm_{N-2-i-j}(p,q)w_{i}[\wh{v}_{j}-\wt{v}_{j}
                                           -(p-q+\wh{u}-\wt{u})\wh{\wt{v}}_{j-1}+p\wh{v}_{j-1}-q\wt{v}_{j-1}]\label{Lpq}\nn\\
\fl                  &&
\eea
under independent variation of $u,v$ and $w$.}

\textit{Proof:} Here we take the usual point of view and consider the action to be the sum of the Lagrangians over all $n,m$, i.e.
\be
 S = \sum_{n,m\in Z} \cL_{pq}.
\ee
The discrete Euler-Lagrange equations arise as a consequence of the requirement that $\dl S = 0$. We have
\bea
\fl 0 & = & \dl S\nn\\
\fl   & = & \sum_{n,m\in Z}\biggl\{\frac{(p-q)(\dl\wh{u}-\dl\wt{u})\gm_{N-1}(p,q)}{p-q+\wh{u}-\wt{u}}-\gm_{N-1}(p,q)(\dl\wh{u}-\dl\wt{u})\nn\\
\fl      && \;\;\;\;-\sum_{j=0}^{N-2}\gm_{N-2-j}(p,q)\bigl\{(\dl\wh{u}-\dl\wt{u})\wh{\wt{v}}_{j}+(\wh{u}-\wt{u})\dl\wh{\wt{v}}_{j}\bigr\}\nn\\
\fl      && -\sum_{i=0}^{N-3}\sum_{j=1}^{N-2-i}\gm_{N-2-i-j}(p,q)\bigl\{\dl w_{i}[\wh{v}_{j}-\wt{v}_{j}
                                           -(p-q+\wh{u}-\wt{u})\wh{\wt{v}}_{j-1}+p\wh{v}_{j-1}-q\wt{v}_{j-1}]\nn\\
\fl      && \;\;\;\;+w_{i}[\dl\wh{v}_{j}-\dl\wt{v}_{j}-(\dl\wh{u}-\dl\wt{u})\wh{\wt{v}}_{j-1}
                           -(p-q+\wh{u}-\wt{u})\dl\wh{\wt{v}}_{j-1}+p\dl\wh{v}_{j-1}-q\dl\wt{v}_{j-1}]\bigr\}\biggr\}\nn\\
\fl   & = & \sum_{n,m\in Z}\biggl\{\biggl(\frac{(p-q)\gm_{N-1}(p,q)}{p-q+\wh{\wt{u}}-\wt{\wt{u}}}
                                         -\frac{(p-q)\gm_{N-1}(p,q)}{p-q+\wh{\wh{u}}-\wh{\wt{u}}}\nn\\
\fl      && \;\;\;\;\;\;\;\;\;\;\;\;\;\; +\sum_{i=0}^{N-3}\sum_{j=0}^{N-3-i}\gm_{N-3-i-j}(p,q)(\wh{\wt{\wt{v}}}_{j}\wt{w}_{i}
                                                                                              -\wh{\wh{\wt{v}}}_{j}\wh{w}_{i})\nn\\
\fl      && \;\;\;\;\;\;\;\;\;\;\;\;\;\; -\sum_{j=0}^{N-3}\gm_{N-2-j}(p,q)(\wh{\wt{\wt{v}}}_{j}-\wh{\wh{\wt{v}}}_{j}-\wt{w}_{j}+\wh{w}_{j})\nn\\
\fl      && \;\;\;\;\;\;\;\;\;\;\;\;\;\; -\wh{\wt{\wt{v}}}_{N-2}+\wh{\wh{\wt{v}}}_{N-2}+\wt{w}_{N-2}-\wh{w}_{N-2}\biggr)\dl\wh{\wt{u}}\nn\\
\fl      && -\sum_{i=0}^{N-3}\sum_{j=0}^{N-3-i}\gm_{N-3-i-j}(p,q)[\wh{v}_{j+1}-\wt{v}_{j+1}
                                                                 -(p-q+\wh{u}-\wt{u})\wh{\wt{v}}_{j}+p\wh{v}_{j}-q\wt{v}_{j}]\dl w_{i}\nn\\
\fl      && +\sum_{i=0}^{N-3}\sum_{j=0}^{N-3-i}\gm_{N-3-i-j}(p,q)[\wh{w}_{i+1}-\wt{w}_{i+1}
                                                                 +(p-q+\wh{u}-\wt{u})w_{i}+q\wh{w}_{i}-p\wt{w}_{i}]\dl\wh{\wt{v}}_{j}\biggr\}.\nn\\
\fl      &&
\eea
The coefficient of $\dl w_{i}$ must be zero, which gives us equation \eqref{v}, the coefficient of $\dl\wh{\wt{v}}_{j}$ must be zero, which gives us equation \eqref{w}, and the remaining term which multiplies $\dl\wh{\wt{u}}$ is two shifted copies of equation \eqref{N-2_better}. Hence the lattice Gel'fand-Dikii system of equations solve the Euler-Lagrange equations for the Lagrangian \eqref{Lpq}.
$\blacksquare$

\paragraph{}
At this point we would like to make several remarks.
\begin{enumerate}
\item For clarity we have chosen to label the Lagrangian with the lattice parameters $p,q$ to indicate it is defined on a plaquette in a 2-dimensional surface corresponding to the respective lattice directions.
\item The Lagrangian \eqref{Lpq} is antisymmetric with respect to the interchange of the lattice directions associated with the parameters $p,q$, a property which will be important when we come to define the multiform structure.
\item Total derivative terms have been included which at first sight may appear superfluous. They do, however, prove necessary in the verification of the closure relation below. For the same reason, we do not have the freedom to multiply the Lagrangian by constants involving the lattice parameters $p,q$, we may only multiply by true constants.
\end{enumerate}

\paragraph{}
The main result of the paper is the following.

\paragraph{}
\noindent
\textbf{Proposition 2:} \emph{The Lagrangian defined by \eqref{Lpq} satisfies the following closure relation on solutions to the lattice GD hierarchy equations when embedded in a 3-dimensional lattice.
\be\label{2dclospq}
 \Dl_{p}\cL_{qr}+\Dl_{q}\cL_{rp}+\Dl_{r}\cL_{pq} = 0,
\ee
where the difference operator $\Dl_{r}$ acts on functions $f$ of $u=u(n_{p},n_{q},n_{r})$ by the formula $\Dl_{r}f(u)=f(\ol{u})-f(u)$, and on a function $g$ of $u$ and its shifts by the formula $\Dl_{r}g(u,\wt{u},\wh{u},\wh{\wt{u}})=g(\ol{u},\wt{\ol{u}},\wh{\ol{u}},\wh{\wt{\ol{u}}})-g(u,\wt{u},\wh{u},\wh{\wt{u}})$.}
\paragraph{}
\textit{Proof:} Firstly, on equation \eqref{v} is is clear that the last term in the Lagrangian will disappear. This leaves us with
\bea
\fl \Gamma & \equiv & \ol{\cL}_{pq}+\wt{\cL}_{qr}+\wh{\cL}_{rp}-\cL_{pq}-\cL_{qr}-\cL_{rp}\nn\\
\fl & = & (p-q)\gm_{N-1}(p,q)\ln\biggl(\frac{p-q+\wh{\ol{u}}-\wt{\ol{u}}}{p-q+\wh{u}-\wt{u}}\biggr)
          +(q-r)\gm_{N-1}(q,r)\ln\biggl(\frac{q-r+\wt{\ol{u}}-\wh{\wt{u}}}{q-r+\ol{u}-\wh{u}}\biggr)\nn\\
\fl    && +(r-p)\gm_{N-1}(r,p)\ln\biggl(\frac{r-p+\wh{\wt{u}}-\wh{\ol{u}}}{r-p+\wt{u}-\ol{u}}\biggr)
          -\gm_{N-1}(p,q)(\wh{\ol{u}}-\wt{\ol{u}}-\wh{u}+\wt{u})\nn\\
\fl    && -\gm_{N-1}(q,r)(\wt{\ol{u}}-\wh{\wt{u}}-\ol{u}+\wh{u})-\gm_{N-1}(r,p)(\wh{\wt{u}}-\wh{\ol{u}}-\wt{u}+\ol{u})\nn\\
\fl    && -\sum_{j=0}^{N-2}\biggl\{\gm_{N-2-j}(p,q)[(\wh{\ol{u}}-\wt{\ol{u}})\wh{\wt{\ol{v}}}_{j}-(\wh{u}-\wt{u})\wh{\wt{v}}_{j}]
                                  +\gm_{N-2-j}(q,r)[(\wt{\ol{u}}-\wh{\wt{u}})\wh{\wt{\ol{v}}}_{j}-(\ol{u}-\wh{u})\wh{\ol{v}}_{j}]\nn\\
\fl    && \;\;\;\;\;\;\;\;\;\;\; +\gm_{N-2-j}(r,p)[(\wh{\wt{u}}-\wh{\ol{u}})\wh{\wt{\ol{v}}}_{j}-(\wt{u}-\ol{u})\wt{\ol{v}}_{j}]\biggr\},
\eea
which, on rearranging the terms, is
\bea
\fl \Gamma & = & -(-p)^{N}\ln\biggl(\biggl(\frac{p-q+\wh{\ol{u}}-\wt{\ol{u}}}{p-q+\wh{u}-\wt{u}}\biggr)
                                    \biggl(\frac{r-p+\wt{u}-\ol{u}}{r-p+\wh{\wt{u}}-\wh{\ol{u}}}\biggr)\biggr)\nn\\
\fl           && -(-q)^{N}\ln\biggl(\biggl(\frac{q-r+\wt{\ol{u}}-\wh{\wt{u}}}{q-r+\ol{u}-\wh{u}}\biggr)
                                    \biggl(\frac{p-q+\wh{u}-\wt{u}}{p-q+\wh{\ol{u}}-\wt{\ol{u}}}\biggr)\biggr)\nn\\
\fl           && -(-r)^{N}\ln\biggl(\biggl(\frac{r-p+\wh{\wt{u}}-\wh{\ol{u}}}{r-p+\wt{u}-\ol{u}}\biggr)
                                    \biggl(\frac{q-r+\ol{u}-\wh{u}}{q-r+\wt{\ol{u}}-\wh{\wt{u}}}\biggr)\biggr)
                 -\gm_{N-1}(p,q)(\wh{\ol{u}}-\wt{\ol{u}}-\wh{u}+\wt{u})\nn\\
\fl           && -\gm_{N-1}(q,r)(\wt{\ol{u}}-\wh{\wt{u}}-\ol{u}+\wh{u})-\gm_{N-1}(r,p)(\wh{\wt{u}}-\wh{\ol{u}}-\wt{u}+\ol{u})
                 -(\wh{\ol{u}}-\wt{\ol{u}})\wh{\wt{\ol{v}}}_{N-2}\nn\\
\fl           && +(\wh{u}-\wt{u})\wh{\wt{v}}_{N-2}
                 -(\wt{\ol{u}}-\wh{\wt{u}})\wh{\wt{\ol{v}}}_{N-2}+(\ol{u}-\wh{u})\wh{\ol{v}}_{N-2}
                 -(\wh{\wt{u}}-\wh{\ol{u}})\wh{\wt{\ol{v}}}_{N-2}+(\wt{u}-\ol{u})\wt{\ol{v}}_{N-2}\nn\\
\fl           && -\sum_{j=0}^{N-3}\biggl\{\gm_{N-2-j}(p,q)[(\wh{\ol{u}}-\wt{\ol{u}})\wh{\wt{\ol{v}}}_{j}-(\wh{u}-\wt{u})\wh{\wt{v}}_{j}]
                                         +\gm_{N-2-j}(q,r)[(\wt{\ol{u}}-\wh{\wt{u}})\wh{\wt{\ol{v}}}_{j}-(\ol{u}-\wh{u})\wh{\ol{v}}_{j}]\nn\\
\fl           && \;\;\;\;\;\;\;\;\;\;\;  +\gm_{N-2-j}(r,p)[(\wh{\wt{u}}-\wh{\ol{u}})\wh{\wt{\ol{v}}}_{j}-(\wt{u}-\ol{u})\wt{\ol{v}}_{j}]\biggr\}.
\eea
We have already shown in section 2 that the lattice KP equation \eqref{LKP} holds provided that equation \eqref{v} holds. Using this fact, it is clear that the logarithm terms disappear. So we are left with
\bea
\fl \Gamma &  = & -\gm_{N-1}(p,q)(\wh{\ol{u}}-\wt{\ol{u}}-\wh{u}+\wt{u})-\gm_{N-1}(q,r)(\wt{\ol{u}}-\wh{\wt{u}}-\ol{u}+\wh{u})
                  -\gm_{N-1}(r,p)(\wh{\wt{u}}-\wh{\ol{u}}-\wt{u}+\ol{u})\nn\\
\fl    && +\ol{u}(\wh{\ol{v}}_{N-2}-\wt{\ol{v}}_{N-2})+\wt{u}(\wt{\ol{v}}_{N-2}-\wh{\wt{v}}_{N-2})
          +\wh{u}(\wh{\wt{v}}_{N-2}-\wh{\ol{v}}_{N-2})\nn\\
\fl    && -\sum_{j=0}^{N-3}\biggl\{\gm_{N-2-j}(p,q)[(\wh{\ol{u}}-\wt{\ol{u}})\wh{\wt{\ol{v}}}_{j}-(\wh{u}-\wt{u})\wh{\wt{v}}_{j}]
                                  +\gm_{N-2-j}(q,r)[(\wt{\ol{u}}-\wh{\wt{u}})\wh{\wt{\ol{v}}}_{j}-(\ol{u}-\wh{u})\wh{\ol{v}}_{j}]\nn\\
\fl    && \;\;\;\;\;\;\;\;\;\;\; +\gm_{N-2-j}(r,p)[(\wh{\wt{u}}-\wh{\ol{u}})\wh{\wt{\ol{v}}}_{j}-(\wt{u}-\ol{u})\wt{\ol{v}}_{j}]\biggr\}.
\eea
Here it is helpful to introduce a new object
\be\label{ep}
 \ep_{j} \equiv \frac{1}{p-q}\bigl(\gm_{j+1}(q,r)-\gm_{j+1}(r,p)\bigr),
\ee
which is invariant under cyclic permutations of $p,q,r$. This is not immediately apparent, but is due to the fact that $\gm_{j}$ obeys the relation
\be
 (p-q)\gm_{j}(p,q)+(q-r)\gm_{j}(q,r)+(r-p)\gm_{j}(r,p) = 0,
\ee
which allows us to write
\bea
\fl \ep_{j} & = & \frac{1}{p-q}\biggl(\gm_{j+1}(q,r)-\gm_{j+1}(r,p)\biggr)\nn\\
\fl         & = & \frac{1}{p-q}\biggl(\gm_{j+1}(q,r)+\frac{1}{r-p}\bigl((p-q)\gm_{j+1}(p,q)+(q-r)\gm_{j+1}(q,r)\bigr)\biggr)\nn\\
\fl         & = & \frac{1}{r-p}\biggl(\gm_{j+1}(p,q)-\gm_{j+1}(q,r)\biggr),
\eea
and this is clearly \eqref{ep} after a cyclic permutation of $p,q$ and $r$. Hence $\ep_{j}$ is invariant under such cyclic permutations. 
The following identity for $\ep_{j}$ also holds
\be\label{ep_identity}
 \ep_{j+1}+r\ep_{j} = \gm_{j+1}(p,q),
\ee
since
\bea
\fl \gm_{j+1}(q,r)+r\gm_{j}(q,r) & = & \frac{(-q)^{j+2}-(-r)^{j+2}}{-q+r}+r\frac{(-q)^{j+1}-(-r)^{j+1}}{-q+r}\nn\\
\fl & = & (-q)^{j+1},
\eea
and similarly $\gm_{j+1}(r,p)+r\gm_{j}(p,q) = (-p)^{j+1}$, so that
\bea
\fl \ep_{j+1}+r\ep_{j} & = & \frac{1}{p-q}\biggl(\gm_{j+2}(q,r)-\gm_{j+2}(r,p)\biggr)+\frac{r}{p-q}\biggl(\gm_{j+1}(q,r)-\gm_{j+1}(r,p)\biggr)\nn\\
 & = & \frac{1}{p-q}\biggl(\bigl(\gm_{j+2}(q,r)+r\gm_{j+1}(q,r)\bigr)-\bigl(\gm_{j+2}(r,p)+r\gm_{j+1}(r,p)\bigr)\biggr)\nn\\
 & = & \frac{(-q)^{j+2}-(-p)^{j+2}}{p-q}\nn\\
 & = & \gm_{j+1}(p,q).
\eea
From the definition, $\ep_{0}=1$, and so we can write 
\bea
\fl \Gamma &  = & -\gm_{N-1}(p,q)(\wh{\ol{u}}-\wt{\ol{u}}-\wh{u}+\wt{u})-\gm_{N-1}(q,r)(\wt{\ol{u}}-\wh{\wt{u}}-\ol{u}+\wh{u})
                  -\gm_{N-1}(r,p)(\wh{\wt{u}}-\wh{\ol{u}}-\wt{u}+\ol{u})\nn\\
\fl    && +\ep_{0}\ol{u}(\wh{\ol{v}}_{N-2}-\wt{\ol{v}}_{N-2})+\ep_{0}\wt{u}(\wt{\ol{v}}_{N-2}-\wh{\wt{v}}_{N-2})
          +\ep_{0}\wh{u}(\wh{\wt{v}}_{N-2}-\wh{\ol{v}}_{N-2})\nn\\
\fl    && -\sum_{j=0}^{N-3}\biggl\{\gm_{N-2-j}(p,q)[(\wh{\ol{u}}-\wt{\ol{u}})\wh{\wt{\ol{v}}}_{j}-(\wh{u}-\wt{u})\wh{\wt{v}}_{j}]
                                  +\gm_{N-2-j}(q,r)[(\wt{\ol{u}}-\wh{\wt{u}})\wh{\wt{\ol{v}}}_{j}-(\ol{u}-\wh{u})\wh{\ol{v}}_{j}]\nn\\
\fl    && \;\;\;\;\;\;\;\;\;\;\; +\gm_{N-2-j}(r,p)[(\wh{\wt{u}}-\wh{\ol{u}})\wh{\wt{\ol{v}}}_{j}-(\wt{u}-\ol{u})\wt{\ol{v}}_{j}]\biggr\}.
\eea
For any $0\leq k\leq N-3$, the expression
\bea
\fl \Lm_k & \equiv &  \ep_{k}\ol{u}(\wh{\ol{v}}_{N-2-k}-\wt{\ol{v}}_{N-2-k})+\ep_{k}\wt{u}(\wt{\ol{v}}_{N-2-k}-\wh{\wt{v}}_{N-2-k})
                     +\ep_{k}\wh{u}(\wh{\wt{v}}_{N-2-k}-\wh{\ol{v}}_{N-2-k})\nn\\
\fl               && -\sum_{j=0}^{N-3-k}\biggl\{\gm_{N-2-j}(p,q)[(\wh{\ol{u}}-\wt{\ol{u}})\wh{\wt{\ol{v}}}_{j}-(\wh{u}-\wt{u})\wh{\wt{v}}_{j}]
                                               +\gm_{N-2-j}(q,r)[(\wt{\ol{u}}-\wh{\wt{u}})\wh{\wt{\ol{v}}}_{j}-(\ol{u}-\wh{u})\wh{\ol{v}}_{j}]\nn\\
\fl               && \;\;\;\;\;\;\;\;\;\;\;    +\gm_{N-2-j}(r,p)[(\wh{\wt{u}}-\wh{\ol{u}})\wh{\wt{\ol{v}}}_{j}-(\wt{u}-\ol{u})\wt{\ol{v}}_{j}]\biggr\}
\eea
can be written as
\bea
\fl \Lm_k & = & \ep_{k}\ol{u}((p-q+\wh{\ol{u}}-\wt{\ol{u}})\wh{\wt{\ol{v}}}_{N-3-k}-p\wh{\ol{v}}_{N-3-k}+q\wt{\ol{v}}_{N-3-k})\nn\\
\fl    && +\ep_{k}\wt{u}((q-r+\wt{\ol{u}}-\wh{\wt{u}})\wh{\wt{\ol{v}}}_{N-3-k}-q\wt{\ol{v}}_{N-3-k}+r\wh{\wt{v}}_{N-3-k})\nn\\
\fl    && +\ep_{k}\wh{u}((r-p+\wh{\wt{u}}-\wh{\ol{u}})\wh{\wt{\ol{v}}}_{N-3-k}-r\wh{\wt{v}}_{N-3-k}+p\wh{\ol{v}}_{N-3-k})\nn\\
\fl    && -\gm_{k+1}(p,q)[(\wh{\ol{u}}-\wt{\ol{u}})\wh{\wt{\ol{v}}}_{N-3-k}-(\wh{u}-\wt{u})\wh{\wt{v}}_{N-3-k}]\nn\\
\fl    && -\gm_{k+1}(q,r)[(\wt{\ol{u}}-\wh{\wt{u}})\wh{\wt{\ol{v}}}_{N-3-k}-(\ol{u}-\wh{u})\wh{\ol{v}}_{N-3-k}]\nn\\
\fl    && -\gm_{k+1}(r,p)[(\wh{\wt{u}}-\wh{\ol{u}})\wh{\wt{\ol{v}}}_{N-3-k}-(\wt{u}-\ol{u})\wt{\ol{v}}_{N-3-k}]\nn\\
\fl    && -\sum_{j=0}^{N-4-k}\biggl\{\gm_{N-2-j}(p,q)[(\wh{\ol{u}}-\wt{\ol{u}})\wh{\wt{\ol{v}}}_{j}-(\wh{u}-\wt{u})\wh{\wt{v}}_{j}]
                                  +\gm_{N-2-j}(q,r)[(\wt{\ol{u}}-\wh{\wt{u}})\wh{\wt{\ol{v}}}_{j}-(\ol{u}-\wh{u})\wh{\ol{v}}_{j}]\nn\\
\fl    && \;\;\;\;\;\;\;\;\;\;\; +\gm_{N-2-j}(r,p)[(\wh{\wt{u}}-\wh{\ol{u}})\wh{\wt{\ol{v}}}_{j}-(\wt{u}-\ol{u})\wt{\ol{v}}_{j}]\biggr\},
\eea
where we have made use of \eqref{v} to eliminate the terms involving shifts of $v_{N-2-k}$. On rearranging,
\bea
\fl \Lm_k & = & [\ep_{k}\ol{u}(p-q+\wh{\ol{u}}-\wt{\ol{u}})+\ep_{k}\wt{u}(q-r+\wt{\ol{u}}-\wh{\wt{u}})
                +\ep_{k}\wh{u}(r-p+\wh{\wt{u}}-\wh{\ol{u}})\nn\\
\fl    && \;\;\;-\gm_{k+1}(p,q)(\wh{\ol{u}}-\wt{\ol{u}})-\gm_{k+1}(q,r)(\wt{\ol{u}}-\wh{\wt{u}})
                -\gm_{k+1}(r,p)(\wh{\wt{u}}-\wh{\ol{u}})]\wh{\wt{\ol{v}}}_{N-3-k}\nn\\
\fl    && +\wt{u}[(-q\ep_{k}+\gm_{k+1}(r,p))\wt{\ol{v}}_{N-3-k}-(-r\ep_{k}+\gm_{k+1}(p,q))\wh{\wt{v}}_{N-3-k}]\nn\\
\fl    && +\wh{u}[(-r\ep_{k}+\gm_{k+1}(p,q))\wh{\wt{v}}_{N-3-k}-(-p\ep_{k}+\gm_{k+1}(q,r))\wh{\ol{v}}_{N-3-k}]\nn\\
\fl    && +\ol{u}[(-p\ep_{k}+\gm_{k+1}(q,r))\wh{\ol{v}}_{N-3-k}-(-q\ep_{k}+\gm_{k+1}(r,p))\wt{\ol{v}}_{N-3-k}]\nn\\
\fl    && -\sum_{j=0}^{N-4-k}\biggl\{\gm_{N-2-j}(p,q)[(\wh{\ol{u}}-\wt{\ol{u}})\wh{\wt{\ol{v}}}_{j}-(\wh{u}-\wt{u})\wh{\wt{v}}_{j}]
                                  +\gm_{N-2-j}(q,r)[(\wt{\ol{u}}-\wh{\wt{u}})\wh{\wt{\ol{v}}}_{j}-(\ol{u}-\wh{u})\wh{\ol{v}}_{j}]\nn\\
\fl    && \;\;\;\;\;\;\;\;\;\;\; +\gm_{N-2-j}(r,p)[(\wh{\wt{u}}-\wh{\ol{u}})\wh{\wt{\ol{v}}}_{j}-(\wt{u}-\ol{u})\wt{\ol{v}}_{j}]\biggr\},
\eea
and then we can use \eqref{LKP} on the very top line to give
\bea
\fl \Lm_k & = & [(r\ep_{k}-\gm_{k+1}(p,q))(\wh{\ol{u}}-\wt{\ol{u}})+(p\ep_{k}-\gm_{k+1}(q,r))(\wt{\ol{u}}-\wh{\wt{u}})\nn\\
\fl          && \;\;\; +(p\ep_{k}-\gm_{k+1}(r,p))(\wh{\wt{u}}-\wh{\ol{u}})]\wh{\wt{\ol{v}}}_{N-3-k}\nn\\
\fl          && +\wt{u}[(-q\ep_{k}+\gm_{k+1}(r,p))\wt{\ol{v}}_{N-3-k}-(-r\ep_{k}+\gm_{k+1}(p,q))\wh{\wt{v}}_{N-3-k}]\nn\\
\fl          && +\wh{u}[(-r\ep_{k}+\gm_{k+1}(p,q))\wh{\wt{v}}_{N-3-k}-(-p\ep_{k}+\gm_{k+1}(q,r))\wh{\ol{v}}_{N-3-k}]\nn\\
\fl          && +\ol{u}[(-p\ep_{k}+\gm_{k+1}(q,r))\wh{\ol{v}}_{N-3-k}-(-q\ep_{k}+\gm_{k+1}(r,p))\wt{\ol{v}}_{N-3-k}]\nn\\
\fl          && -\sum_{j=0}^{N-4-k}\biggl\{\gm_{N-2-j}(p,q)[(\wh{\ol{u}}-\wt{\ol{u}})\wh{\wt{\ol{v}}}_{j}-(\wh{u}-\wt{u})\wh{\wt{v}}_{j}]
                                          +\gm_{N-2-j}(q,r)[(\wt{\ol{u}}-\wh{\wt{u}})\wh{\wt{\ol{v}}}_{j}-(\ol{u}-\wh{u})\wh{\ol{v}}_{j}]\nn\\
\fl          && \;\;\;\;\;\;\;\;\;\;\;    +\gm_{N-2-j}(r,p)[(\wh{\wt{u}}-\wh{\ol{u}})\wh{\wt{\ol{v}}}_{j}-(\wt{u}-\ol{u})\wt{\ol{v}}_{j}]\biggr\}.
\eea
Here the identity \eqref{ep_identity} comes into play, to give
\bea
\fl \Lm_k & = & \ep_{k+1}\ol{u}(\wh{\ol{v}}_{N-3-k}-\wt{\ol{v}}_{N-3-k})+\ep_{k+1}\wt{u}(\wt{\ol{v}}_{N-3-k}-\wh{\wt{v}}_{N-3-k})
               +\ep_{k+1}\wh{u}(\wh{\wt{v}}_{N-3-k}-\wh{\ol{v}}_{N-3-k})\nn\\
\fl          && -\sum_{j=0}^{N-4-k}\biggl\{\gm_{N-2-j}(p,q)[(\wh{\ol{u}}-\wt{\ol{u}})\wh{\wt{\ol{v}}}_{j}-(\wh{u}-\wt{u})\wh{\wt{v}}_{j}]
                                          +\gm_{N-2-j}(q,r)[(\wt{\ol{u}}-\wh{\wt{u}})\wh{\wt{\ol{v}}}_{j}-(\ol{u}-\wh{u})\wh{\ol{v}}_{j}]\nn\\
\fl          && \;\;\;\;\;\;\;\;\;\;\; +\gm_{N-2-j}(r,p)[(\wh{\wt{u}}-\wh{\ol{u}})\wh{\wt{\ol{v}}}_{j}-(\wt{u}-\ol{u})\wt{\ol{v}}_{j}]\biggr\}\nn\\
\fl       & = & \Lm_{k+1},
\eea
which means that for any $0\leq j,k \leq N-2$ we have $\Lm_j = \Lm_k$. This allows us to greatly simplify $\Gamma$, since
\bea
\fl \Gamma &  = & -\gm_{N-1}(p,q)(\wh{\ol{u}}-\wt{\ol{u}}-\wh{u}+\wt{u})-\gm_{N-1}(q,r)(\wt{\ol{u}}-\wh{\wt{u}}-\ol{u}+\wh{u})
                  -\gm_{N-1}(r,p)(\wh{\wt{u}}-\wh{\ol{u}}-\wt{u}+\ol{u})\nn\\
\fl            && +\Lm_{0}\nn\\
\fl        &  = & -\gm_{N-1}(p,q)(\wh{\ol{u}}-\wt{\ol{u}}-\wh{u}+\wt{u})-\gm_{N-1}(q,r)(\wt{\ol{u}}-\wh{\wt{u}}-\ol{u}+\wh{u})
                  -\gm_{N-1}(r,p)(\wh{\wt{u}}-\wh{\ol{u}}-\wt{u}+\ol{u})\nn\\
\fl            && +\Lm_{N-2}\nn\\
\fl         & = & -\gm_{N-1}(p,q)(\wh{\ol{u}}-\wt{\ol{u}}-\wh{u}+\wt{u})-\gm_{N-1}(q,r)(\wt{\ol{u}}-\wh{\wt{u}}-\ol{u}+\wh{u})
                  -\gm_{N-1}(r,p)(\wh{\wt{u}}-\wh{\ol{u}}-\wt{u}+\ol{u})\nn\\
\fl            && +\ep_{N-2}\ol{u}(\wh{\ol{u}}-\wt{\ol{u}})+\ep_{N-2}\wt{u}(\wt{\ol{u}}-\wh{\wt{u}})
                  +\ep_{N-2}\wh{u}(\wh{\wt{u}}-\wh{\ol{u}}).
\eea
Using once again the equations \eqref{LKP} and then \eqref{ep_identity}, this is 
\bea
\fl \Gamma &  = & -\gm_{N-1}(p,q)(\wh{\ol{u}}-\wt{\ol{u}}-\wh{u}+\wt{u})-\gm_{N-1}(q,r)(\wt{\ol{u}}-\wh{\wt{u}}-\ol{u}+\wh{u})
                  -\gm_{N-1}(r,p)(\wh{\wt{u}}-\wh{\ol{u}}-\wt{u}+\ol{u})\nn\\
\fl            && +r\ep_{N-2}(\wh{\ol{u}}-\wt{\ol{u}}-\wh{u}+\wt{u})+p\ep_{N-2}(\wt{\ol{u}}-\wh{\wt{u}}-\ol{u}+\wh{u})
                  +q\ep_{N-2}(\wh{\wt{u}}-\wh{\ol{u}}-\wt{u}+\ol{u})\nn\\
\fl         & = &  (r\ep_{N-2}-\gm_{N-1}(p,q))(\wh{\ol{u}}-\wt{\ol{u}}-\wh{u}+\wt{u})
                  +(p\ep_{N-2}-\gm_{N-1}(q,r))(\wt{\ol{u}}-\wh{\wt{u}}-\ol{u}+\wh{u})\nn\\
\fl            && +(q\ep_{N-2}-\gm_{N-1}(r,p))(\wh{\wt{u}}-\wh{\ol{u}}-\wt{u}+\ol{u})\nn\\
\fl         & = & -\ep_{N-1}(\wh{\ol{u}}-\wt{\ol{u}}-\wh{u}+\wt{u})-\ep_{N-1}(\wt{\ol{u}}-\wh{\wt{u}}-\ol{u}+\wh{u})
                  -\ep_{N-1}(\wh{\wt{u}}-\wh{\ol{u}}-\wt{u}+\ol{u})\nn\\
\fl         & = & 0.
\eea
Thus the closure relation is verified. $\blacksquare$

\paragraph{}
Note that in the above computation, the only equations used were the equation involving the $v_{j}$  \eqref{v}, and the lattice KP equation \eqref{LKP}, the latter was shown earlier to be a consequence of copies of \eqref{v}.

\section{Lattice Boussinesq and KdV equations}
The lattice Boussinesq equation deserves special mention as it has attracted much interest lately, for example with regard to the Pentagram map \cite{OvsSchTab2009}. It is a particular case of the lattice GD hierarchy, taking $N=3$, and as such can be written as a system of equations in the variables $u,v_{1},w_{1}$. However, it is possible to eliminate $v_{1},w_{1}$ and express the equation in terms of the variable $u$ only, as follows.
\bea
\fl \frac{p^3-q^3}{p-q+\wh{\wt{u}}-\wt{\wt{u}}}-\frac{p^3-q^3}{p-q+\wh{\wh{u}}-\wh{\wt{u}}}
+(p+2q)(\wh{\wt{\wt{u}}}+\wh{u})-(2p+q)(\wh{\wh{\wt{u}}}+\wt{u})&&\nn\\
\fl \;\;\;\;\;\;\;\;\;\;\;\;\;\;\;\;\;\;\;\;\;\;\;\;
+(p-q+\wh{\wh{\wt{u}}}-\wh{\wt{\wt{u}}})\wh{\wh{\wt{\wt{u}}}}+(p-q+\wh{u}-\wt{u})u+\wt{u}\wh{\wt{\wt{u}}}-\wh{u}\wh{\wh{\wt{u}}} & = & 0.
\eea
Starting from the action given in \cite{NijPapCapQui1992}, we need to make only minor modifications in order to arrive at a Lagrangian in terms of the variable $u$ only which satisfies the closure relation \eqref{2dclospq} on solutions to the lattice KP equation \eqref{LKP},
\bea
\fl \cL_{pq} & = & (p^3-q^3)\ln(p-q+\wh{u}-\wt{u})-(p^2+pq+q^2)(\wh{u}-\wt{u})+(p+q)(\wh{u}-\wt{u})\wh{\wt{u}}\nn\\
\fl             && +(p-q+\wh{u}-\wt{u})u\wh{\wt{u}}-pu\wh{u}+qu\wt{u}.
\eea
That the above Lagrangian satisfies the closure relation \eqref{2dclospq} can easily be verified by direct computation.

\paragraph{}
The lattice KdV equation, which is the member of the lattice GD hierarchy where $N=2$, is a more degenerate case and needs to be treated separately. Here we do not have the equations \eqref{v} and \eqref{w} as $N$ is too small, we have only the equation \eqref{N-2_better}, which is
\be\label{KdV}
(p+q+u-\wh{\wt{u}})(p-q+\wh{u}-\wt{u})=p^2-q^2.
\ee
A Lagrangian was first given in \cite{CapNijPap1991}, which is equivalent to the following Lagrangian
\be
 \cL_{pq} = -(p^2-q^2)\ln(p-q+\wh{u}-\wt{u})+(\wh{u}-\wt{u})(p+q-\wh{\wt{u}}).
\ee
Again, the closure relation holds on solutions of the lattice KP equation \eqref{LKP}, but we need to use copies of equation \eqref{KdV} in 3 lattice directions to show that \eqref{LKP} does indeed hold. The lattice KdV equation is a case already treated in \cite{LobNij2009}.

\section{Lagrangian multiform structure}
Having established a Lagrangian for each member of the lattice GD hierarchy satisfying the closure relation \eqref{2dclospq}, we now interpret this result in terms of a Lagrangian multiform structure. In fact, the existence of the closure relation allows us to develop the variational principle proposed in \cite{LobNij2009} for this class of systems. This comprises the following.

Noting that the Lagrangian \eqref{Lpq} is defined on an elementary plaquette, we can define an action $S$ for any given surface $\sigma$ consisting of a connected configuration of elementary plaquettes $\sigma_{pq}$ in the multidimensional lattice (where the labelling by the lattice parameters $p,q$ indicates $\sigma_{pq}$ lives on the sublattice corresponding to the respective lattice directions) by summing the Lagrangian contributions from each of the plaquettes in the surface $\sigma$, i.e.
\be
 S[u,v_{1},\dots,v_{N-2},w_{1},\dots,w_{N-2};\sigma] = \sum_{\sigma_{pq}\in \sigma}\cL_{pq},
\ee
taking into account the orientation of the plaquette (as noted earlier, $\cL_{pq}$ has the property of antisymmetry with respect to interchange of the two lattice directions, so this sum is well-defined). This action depends not only on the dependent variables, but also on the geometry of the independent variables. The variational principle proposed in \cite{LobNij2009} amounts to the following reasoning. Imposing independence of the action under local variations of the surface, keeping the boundary fixed, requires the closure relation to hold. Furthermore, the surface independence allows us to locally deform the surface in any way we choose away from the boundary. In particular, away from the boundary we may render it locally flat so that we have here a regular 2-dimensional lattice on which we can apply the variational principle leading to the usual discrete Euler-Lagrange equations. It then follows from the proof of Proposition 2 in the previous section that for this specific Lagrangian, the equations of motion \eqref{v},\eqref{w} and \eqref{N-2_better} are compatible with the surface independence. We view this circular mechanism behind the variational principle as a manifestation of multidimensional consistency on the level of the Lagrangian. 

The requirement that the closure relation should hold specifies to some extent the Lagrangian. For all examples in \cite{LobNij2009}, for the closure relation to hold it was necessary for the Lagrangian to be antisymmetric with respect to the interchange of the two lattice directions, although total derivatives could be added so long as they were also antisymmetric. Here, however, it seems we are even more constrained, as in general we are not free to add such total derivatives. That the closure relation places such restrictions on the Lagrangian could be of relevance in the inverse problem of Lagrangian mechanics, a field of study which dates back to the 1880s \cite{Hel1886}, cf. \cite{Ton1985} for a review.

It would also be interesting to see whether a similar closure relation holds for the continuous GD hierarchy. A Lagrangian for the generating partial differential equation (PDE) for the KdV hierarchy \cite{NijHonJos2000} was shown to obey a continuous analogue of the closure relation in \cite{LobNij2009}. The generating PDE for the Boussinesq hierarchy along with its Lagrangian appeared in \cite{TonNij2005i,TonNij2005ii}, it is expected that this would also obey a continuous analogue of the closure relation.

\section*{Acknowledgments}
SBL was supported by the UK Engineering and Physical Sciences Research Council (EPSRC).

\end{document}